%% file: DEmodels.v3.tex
\documentclass[useAMS,usenatbib,usegraphicx]{mn2e}

\usepackage{graphicx}
\usepackage[latin1]{inputenc}
\usepackage{color}
\usepackage{times}
\usepackage{natbib}
\usepackage{setspace}
\newif\ifAMStwofonts
\AMStwofontstrue
\definecolor{red}{rgb}{1,0.,0.}

\newcommand{\munich}{{\sc l-galaxies }}
\newcommand{\lcdm}{$\Lambda$CDM }
\newcommand{\fr}{$f(R)$-gravity }

\def\lesssim{\lower.5ex\hbox{$\; \buildrel < \over \sim \;$}}
\def\gtrsim{\lower.5ex\hbox{$\; \buildrel > \over \sim \;$}}
\title[Galaxy Formation in \fr] {Semi-analytic galaxy formation in
  \fr cosmologies}

\author[Fontanot et al.]{
  \parbox[t]{\textwidth}{Fabio Fontanot$^{1,2}$\thanks{E-mail:
      fabio.fontanot@h-its.org}, Ewald Puchwein$^{1}$, Volker
    Springel$^{1,3}$, Davide Bianchi$^{4,5}$}
    \vspace*{8pt}\\
    $^1$ Heidelberger Institut f\"ur Theoretische Studien (HITS), Schloss-Wolfsbrunnenweg 35, 69118 Heidelberg, Germany\\
    $^2$ Institut f\"ur Theoretische Physik, Philosophenweg 16, 69120, Heidelberg, Germany \\
    $^3$ Zentrum f\"ur Astronomie der Universit\"at Heidelberg, ARI, M\"onchhofstrasse 12-14, 69120 Heidelberg, Germany \\
    $^4$ INAF-Osservatorio Astronomico di Brera, Via Bianchi 46, I-23807 Merate (LC), Italy\\
    $^5$ Dipartimento di Fisica, Universit\`a degli Studi di Milano, via Celoria 16, I-20133 Milano, Italy \\
}

\begin{document}
\date{Accepted ... Received ...}

\maketitle

\begin{abstract} 
  Modifications of the equations of general relativity at large
  distances offer one possibility to explain the observed properties
  of our Universe without invoking a cosmological constant. Numerous
  proposals for such modified gravity cosmologies exist, but often
  their consequences for structure formation in the non-linear sector
  are not yet accurately known.  In this work, we employ
  high-resolution numerical simulations of \fr models coupled with a
  semi-analytic model (SAM) for galaxy formation to obtain detailed
  predictions for the evolution of galaxy properties. The \fr models
  imply the existence of a `fifth-force', which is however locally
  suppressed, preserving the successes of general relativity on solar
  system scales.  We show that dark matter haloes in \fr models are
  characterized by a modified virial scaling with respect to the \lcdm
  scenario, reflecting a higher dark matter velocity dispersion at a
  given mass. This effect is taken into account in the SAM by an
  appropriate modification of the mass--temperature relation. We find
  that the statistical properties predicted for galaxies (such as the
  stellar mass function and the cosmic star formation rate) in \fr
  show generally only very small differences relative to $\Lambda$CDM,
  smaller than the dispersion between the results of different SAM
  models, which can be viewed as a measure of their systematic
  uncertainty.  We also demonstrate that galaxy bias is not able to
  disentangle between \fr and the standard cosmological
  scenario. However, \fr imprints modifications in the linear growth
  rate of cosmic structures at large scale, which can be recovered
  from the statistical properties of large galaxy samples.
\end{abstract}

\begin{keywords}
  galaxies: formation - galaxies: evolution - galaxies:fundamental properties
\end{keywords}

\section{Introduction}\label{sec:intro}

Our knowledge of the basic properties of the Universe has seen
considerable advances in recent decades, culminating in the accurate
determination of the key cosmological parameters \citep[see
  e.g.][]{Planck_cosmpar,Komatsu09}. Measurements of the accelerated
expansion and of the curvature and matter content of the Universe have
led to the conclusion that some unknown form of {\it Dark Energy} (DE,
hereafter) accounts for $\sim 70$ per cent of its energy density
today.  In the standard $\Lambda$CDM model, the dark energy is
described by a classical cosmological constant, and so far, available
experiments have been unable to decide whether the true nature of the
dark energy differs from a cosmological constant, something that is
generally expected on theoretical grounds.  Numerous scenarios,
ranging from scalar field theories (e.g. quintessence) to
modifications of the equations of general relativity, have been
proposed as alternatives to the standard cosmological constant model,
where a homogeneous and static energy field fills the whole Universe
at any cosmic epoch.

Future wide galaxy surveys (like EUCLID, \citealt{Laureijs11}) have
been proposed as powerful experiments to constrain the basic
properties of DE and disentangle different proposed scenarios.
However, the success of these missions hinges on a precise
understanding of the relationship between the physical processes
responsible for galaxy formation and evolution, and the assembly of
the cosmic large-scale structure. While the non-linear evolution of
virialized structures in a variety of cosmological scenarios has been
explored by means of high-resolution $N$-body simulations (see
e.g. \citealt{GrossiSpringel09} for quintessence models;
\citealt{Schmidt09b, KhouryWyman09, LiZhaoKoyama12} for modified
gravity models; \citealt{Baldi12} for coupled DE models), their
influence on the properties of galaxies has not yet been explored in
much detail. This is a question of fundamental importance, since the
envisaged tests for constraining the properties of DE ultimately
depend on our understanding of how galaxies trace the distribution of
dark matter (DM) on large scales.

Semi-analytic models (SAMs, hereafter) of galaxy formation, and
hydrodynamic numerical simulations alike, resort to simplified yet
physically grounded, analytical approximations to describe the
relevant physical mechanisms (such as gas cooling, star formation,
black hole accretion, feedbacks) and their interplay, as a function of
the physical properties of the simulated objects (e.g. stellar mass,
gas content) and/or their environment (e.g. host halo mass). This
approach intrinsically involves a phenomenological parameter space,
which is constrained by means of a calibration procedure against a
selected subset of (mainly) low-redshift observations. The current
generation of models has been shown to successfully reproduce a large
number of observational evidences.  Nevertheless, our detailed
knowledge of the chain of physical processes responsible for galaxy
formation and evolution is still limited, {\it even} for a standard
\lcdm cosmology. A number of well known
discrepancies\footnote{Including, e.g., the evolution of low-mass
  galaxies \citep{Fontanot09b,Weinmann12,Henriques13}; the properties
  of satellite galaxies \citep{Weinmann09,Maccio10,BoylanKolchin12};
  and perhaps to a lesser degree the baryon fractions in galaxy
  clusters \citep{McCarthy07}.} between the predictions of theoretical
models and observational constraints still exist. Due to the
considerable complexity of the coupled non-linear physical processes
of the baryonic gas and due to the significant level of degeneracy
among the relevant parametrizations \citep[see e.g.][]{Henriques09},
this can however not easily be translated into constraints of the
underlying cosmological model.  It is thus a high priority in the
framework of future space missions to identify any modification in
model predictions that can be directly and robustly ascribed to the
effect of a given DE scenario. An important goal is to quantify the
performance of different cosmological tests based on statistical
galaxy properties, marginalised over the uncertainties of galaxy
formation physics.

This paper is the second of a series aimed at studying the impact of
alternative DE cosmologies on the properties of galaxy populations, as
predicted by SAMs. In the first paper \citep[][hereafter Paper
  I]{Fontanot12c}, we considered a specific class of DE cosmologies,
namely the Early Dark Energy (EDE) models, in which the DE constitutes
a small but finite fraction of the total energy density at the time of
matter-radiation equality, leading to an earlier formation of
structures with respect to the \lcdm cosmology at an equal amplitude
of the present-day clustering strength. In this work, we extend the
results presented in Paper I to a new class of DE cosmologies, i.e.~by
modifying the left-hand side of the Einstein equation and invoking new
degrees of freedom. In particular, the addition of a non-linear
function of the Ricci scalar $R$ to the Einstein-Hilbert action can
mimic cosmic acceleration, while the postulated extra degree of
freedom is giving rise to a long-range `fifth force'.  Searches for
`fifth-force' effects and/or violations of the weak equivalence
principle have established tight constraints on the coupling of these
new degrees of freedom with the fields of the standard
cosmology. These models might be split into two classes, according to
the proposed mechanism for suppressing the scalar force in dense
environments: either the field becomes massive such that the Compton
wavelength of the interaction becomes small (such as the `Chameleon'
Effect, \citealt{KhouryWeltman04}), or the coupling to matter becomes
extremely weak (such as the \citealt{Vainshtein72} mechanism or the
`Symmetron' model, \citealt{HinterbichlerKhoury10}). In both cases, we
have a `screening effect' which ensures that the agreement of local
tests with the prediction of standard general relativity is still
preserved.

It has been suggested that one of the cleanest ways to distinguish
between standard general relativity and modified gravity theories lies
in the analysis of the linear growth rate $f(z)$, i.e. by measuring
how rapidly structures are being assembled in the Universe \citep[see
  e.g][]{Guzzo08}. In general, $f(z)$ is related to the matter density
parameter $\Omega_m(z)$ through the equation $f(z) \sim
\Omega_m^\gamma$. Models with similar expansion history $H(z)$ but
different theories of gravity predict a different growth rate $f(z)$
and index $\gamma$. Any deviation of $\gamma$ from the standard value
predicted in general relativity ($-0.55$) would represent strong
evidence in favour of modifications of gravity, rather than pointing
to exotic new ingredients in the particle content of the Universe. A
similar derived quantity is the parameter $E_{\rm G}$ which combines
the structure growth rate with measures of large-scale gravitational
lensing and galaxy clustering \citep{Reyes10}.

This paper is organized as follows. In Section~\ref{sec:models}, we
introduce the cosmological numerical simulations and semi-analytic
models we use in our analysis. We then present the predicted galaxy
properties and compare them among different cosmologies in
Section~\ref{sec:results}. Finally, we discuss our conclusions in
Section~\ref{sec:final}.

\section{Models}\label{sec:models}
\begin{table*}
  \caption{Cosmological parameters of our simulation set. The columns
    list from left to right: total dark matter density, dark energy
    content at $z=0$, present-day expansion rate, power spectrum
    normalization, equation of state parameter, and the quantities
    describing the early dark energy or the $f(R)$ gravity
    modification, respectively.}
  \label{tab:results}
  \renewcommand{\footnoterule}{}
  \centering
  \begin{tabular}{ccccccccc}
    \hline
     & $\Omega_m$ & $\Omega_{\Lambda,0}$ & $h$ & $\sigma_8$ & $w_0$ & $\Omega_{\rm de,e}$ & $\eta$ & $|\bar{f}_{R0}|$ \\
    \hline
    \lcdm & 0.272 & 0.728 & 0.704 & 0.809 & -1.0  & --- &  --- & --- \\
    EDE3  & 0.272 & 0.728 & 0.704 & 0.809 & -0.93 & 2 $\times 10^{-3}$ & --- & --- \\
    FoR1  & 0.272 & 0.728 & 0.704 & 0.809 & -1.0  & --- & 1.0 & $10^{-4}$ \\
    \hline
  \end{tabular}
\end{table*}

\subsection{\fr parametrization}

In this paper, we focus on a particular class of modified gravity
cosmologies, namely \fr models. In particular, we consider the same
class of modified gravity models studied in \citet{Puchwein13}, with a
parametrization first introduced by \citet{HuSawicki07}. We assume a
modification of the Einstein-Hilbert action of the form

\begin{equation}
S = \int {\rm d}^4x\, \sqrt{-g} \left[ \frac{R+f(R)}{16 \pi G} + \mathcal{L}_m
  \right],
\end{equation}
\noindent
where $R$ is the Ricci scalar and $\mathcal{L}_m$ the matter
Lagrangian. \citet{HuSawicki07} argued that the functional form for
$f(R)$ is constrained by high- and low-redshift observational
data, and proposed a general class of broken power law models to parameterize
$f(R)$ in the form
\begin{equation}
f(R) = -m^2 \frac{c_1(R/m^2)^\eta}{c_2 (R/m^2)^{\eta}+1},
\end{equation}
\noindent
where $m^2=H_0^2 \Omega_m$ and $\eta$, $c_1$ and $c_2$ are free
parameters. It is possible to derive relations between these free
parameters by requiring the model to closely reproduce a flat \lcdm
cosmic expansion history, which yields

\begin{equation}
\frac{c_1}{c_2} \approx 6 \frac{\Omega_\Lambda}{\Omega_m}
\end{equation}
\noindent
and

\begin{equation}
f_R = \frac{{\rm d} f(R)}{{\rm d}R} = -\eta \frac{c_1}{c_2^2} \left( \frac{m^2}{R} \right)^{\eta+1}.
\end{equation}
\noindent
It is then possible to completely describe this class of \fr models by
using two parameters $\eta$ and $\bar{f}_{R0}$ (i.e. the background
value of $f_R$ at $z=0$). In this paper, we will consider a \fr model
with $\eta=1$ and $|\bar{f}_{R0}|=10^{-4}$, which corresponds to a
large deviation from a \lcdm cosmology. While this value is already in
tension with local tests of gravity, it is still interesting to
consider it to maximize the impact of \fr models on the statistical
properties of galaxy populations.

\subsection{Numerical Simulations}

In this work we consider a set of numerical $N$-body simulations of DM
only runs (see Table~\ref{tab:results}). We use a \fr run and, for
comparison, also consider two additional cosmological runs, assuming a
standard \lcdm cosmology and an EDE model, respectively. The former
model has been taken from the EDE runs considered in Paper I, and is
named EDE3 following the same labelling convention as in Paper I. In
all simulations, we assume a flat universe with matter density
parameter $\Omega_m=0.272$, Hubble parameter $h=0.704$, Gaussian
density fluctuations with a scale-invariant primordial power spectrum
with spectral index $n=1$, and a normalization of the linearly
extrapolated $z=0$ power spectrum equal to $\sigma_8=0.8$.

We generate initial conditions for all the simulations using the {\sc
  n-genic} code and run the simulation using the cosmological code
{\sc gadget-3} \citep[last described in][]{Springel05c}. For the EDE3
run, we employ the same code version as in \citet{GrossiSpringel09} to
account for EDE cosmologies with a redshift-dependent equation of
state. Finally, we perform the FoR1 run with \fr using the new {\sc
  mg-gadget} code developed by \citet{Puchwein13}.  This code augments
the usual algorithms of {\sc gadget} for ordinary gravity with a
multigrid-accelerated Newton-Gauss-Seidel relaxation solver on an
adaptive mesh to efficiently solve for perturbations in the scalar
degree of freedom of the modified gravity model. All simulations have
been run in periodic boxes $100\,h^{-1}{\rm Mpc}$ on a side, using
$512^3$ particles (corresponding to a mass resolution of $5.62 \times
10^8\,h^{-1}{\rm M}_\odot$). We use initial conditions with the same
phases and mode amplitudes in the \lcdm and \fr runs to ensure a
similar realization of the large scale structure and allow an
object-by-object comparison. It is worth stressing that, at variance
with Paper I, we do not constrain the \lcdm and \fr simulations to
have similar matter power spectrum at $z=0$: instead we require the
initial conditions for these runs to share the same amplitude of the
initial power spectrum at the last scattering surface.

For each run, 64 simulation snapshots were produced at the same output
redshifts used in the Millennium simulation project \citep{Springel05}
and in Paper I.  
Group catalogues have been constructed using the friend-of-friend
(FOF) algorithm with a linking length of 0.2 in units of the mean
particle separation. Each group has then been decomposed into
gravitationally bound substructures using {\sc subfind}
\citep{Springel01}, and the resulting subhalo catalogues were used to
construct merger history trees as explained in detail in
\citet{Springel05}. Only subhaloes that retained at least 20 bound
particles after the gravitational unbinding procedure were kept for
the tree construction.

\subsection{Semi-Analytic Models}
\begin{figure*}
  \centerline{ \includegraphics[width=18cm]{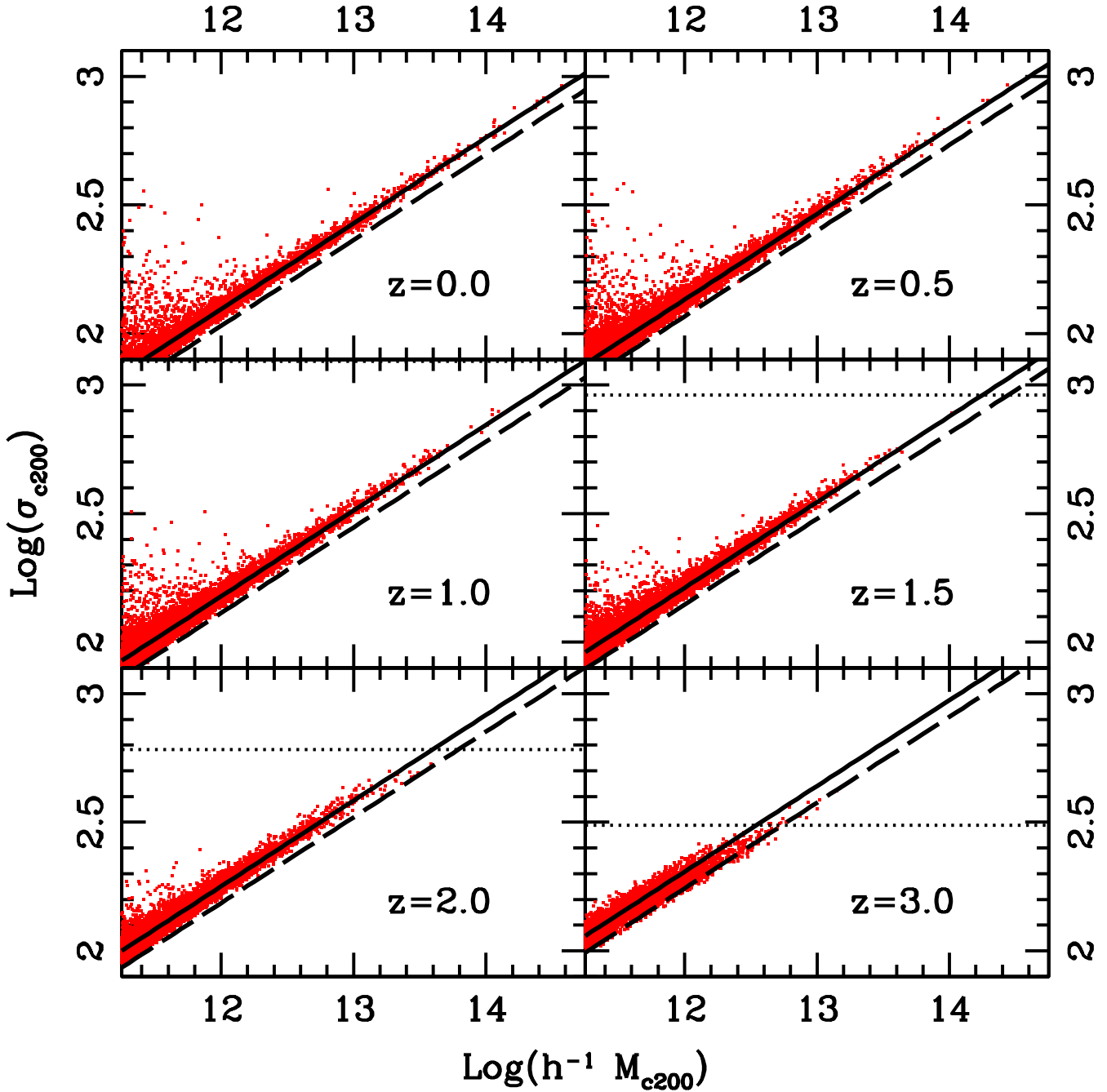} }
  \caption{Virial scaling relation $M_{200}$-$\sigma_{200}$ for the
    haloes in the FoR1 run. Solid and dashed lines show the best fit
    formulae for the relations in FoR1 and \lcdm cosmologies,
    respectively (see text for more details).}\label{fig:virial}
\end{figure*}
\begin{figure*}
  \centerline{ \includegraphics[width=18cm]{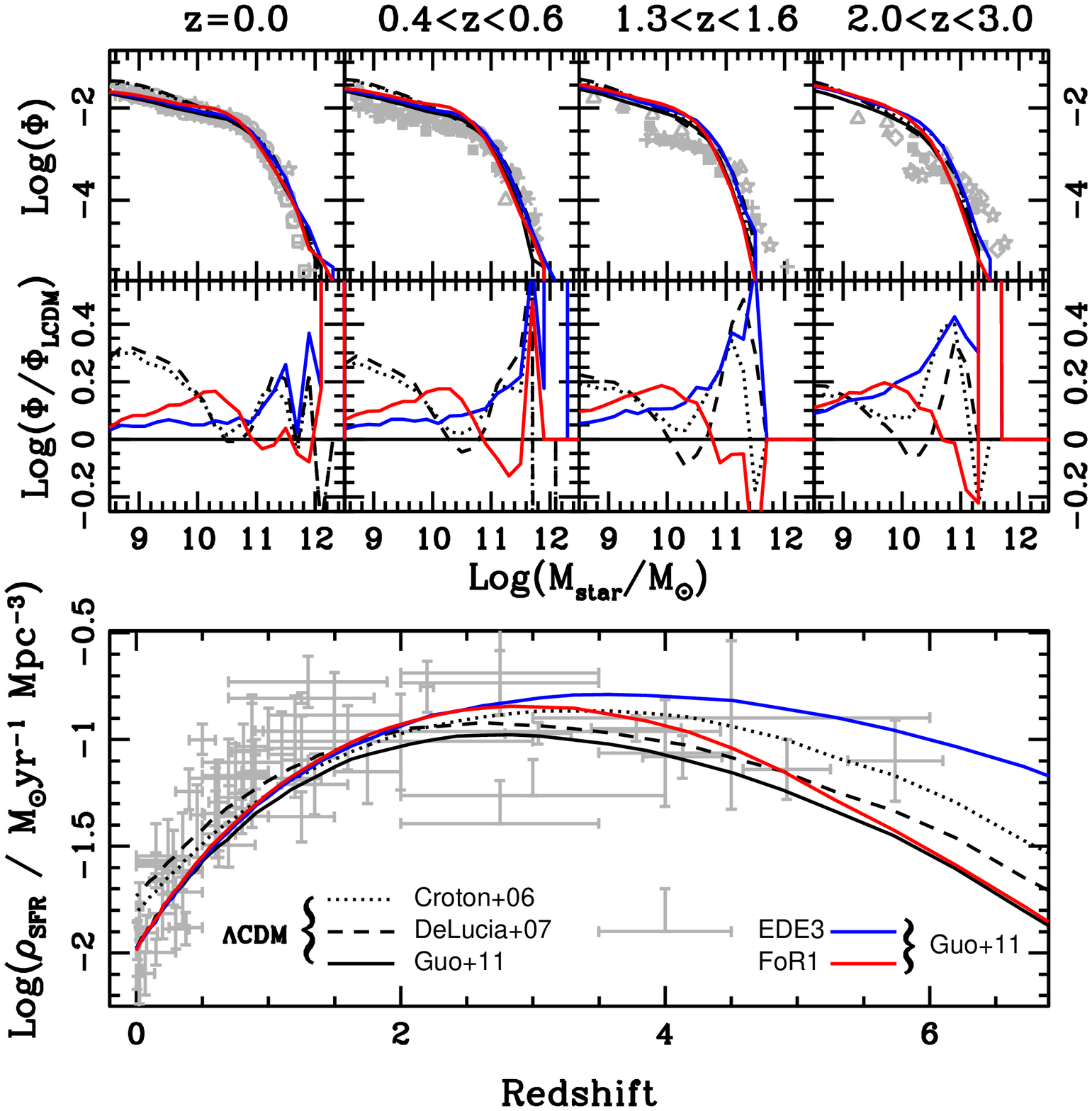} }
  \caption{A collection of SAM predictions in different cosmological
    scenarios. In each panel the solid black/blue/red lines refer to
    our SAM predictions in $\Lambda$CDM, Early Dark Energy (EDE3) and
    \fr (FoR1) cosmologies, respectively. Black dashed and dotted
    lines refer to the predictions of the \citet{DeLuciaBlaizot07} and
    \citet{Croton06} SAMs for the $\Lambda$CDM cosmology,
    respectively. {\it Upper Panel}: redshift evolution of the
    predicted stellar mass function. Grey points refer to the stellar
    mass function compilation from \citet[][see references
      herein]{Fontanot09b}. {\it Lower Panel}: redshift evolution of
    the cosmic star formation rate density. Grey points refer to a
    cosmic star formation density compilation from \citet[][see
      references herein]{Hopkins04}.}\label{fig:gen_sam}
\end{figure*}

In this paper, we use the same approach as proposed in Paper I. We
consider three different versions of {\sc l-galaxies}, based on the
code originally developed by \citet{Springel05}, namely the versions
and corresponding models presented in \citet{Guo11},
\citet{DeLuciaBlaizot07}, and \citet{Croton06}. These models share the
same code structure and were designed to work on Millennium-like
merger trees. They thus represent a coherent set of models that can be
used to study the intra-model variance induced by different choices
for the approximation of galaxy formation physics. A comparison of the
corresponding systematic model uncertainties to the changes in the
predictions when the underlying cosmology is changed is quite
informative for assessing how easily these differences could be
detected. We refer the reader to the original papers for a more
detailed discussion of the modelling of the relevant physical
mechanisms in each model. All SAMs have been calibrated by requiring
them to reproduce a well-defined set of low-redshift reference
observations.  The differences\footnote{The main differences between
  the three models are: (a) the treatment of dynamical friction and
  merger times, the stellar initial mass function and the dust model
  (from \citealt{Croton06} to \citealt{DeLuciaBlaizot07}); (b) the
  modeling of supernovae feedback, the treatment of satellite galaxy
  evolution, tidal stripping and mergers (improved from
  \citealt{DeLuciaBlaizot07} to \citealt{Guo11}).} in the treatment of
physical processes in these SAM versions are large enough to require a
general re-calibration of the main model parameters.

In Paper I, the \citet{Guo11} model has been modified to run
self-consistently on EDE cosmologies. In this work we extend these
modifications to include \fr models. For the remainder of this paper,
whenever we refer to the \citet{Guo11} model, we actually refer to our
modified code. The first change in the code follows closely from Paper
I, as we implement in the SAM a tool to specify the Hubble function
$H(a)$ through an external file containing a user-generated
expression, thus taking into account the possible variation of the
expansion rate connected to a DE cosmology. A second relevant
modification comes from the analysis of the properties of DM haloes in
the FoR1 simulation. In Figure~\ref{fig:virial}, we show the virial
scaling relation between the total DM mass inside a sphere with
interior mean density 200 times the critical density at a given
redshift, $M_{200}(z)$, and the one-dimensional velocity
dispersion $\sigma_{200}$ inside the same radius. \citet{Evrard08}
showed that haloes in a \lcdm cosmological run follow with good
approximation the theoretical relation (dashed line):

\begin{equation}\label{eq:scaling}
\sigma_{200} \propto [ h(z) M_{200}(z) ]^{1/3}.
\end{equation}
\noindent
We find that haloes in \fr are generally offset from this relation,
and only haloes massive enough for the screening effect to be
effective lie on the expected relation for the \lcdm cosmology. A
simple estimate of the halo mass or halo velocity dispersion scale on
which the Chameleon mechanism screens modifications of gravity as a
function of redshift can be obtained by comparing the background value
of the scalar degree of freedom $\bar{f}_R(z)$ to the depth of a halo
potential well. More precisely, screening happens in deep potential
wells roughly when
  
\begin{equation}
|\phi_N| > \frac{3 \, c^2 \, |\bar{f}_R(z)| }{2} ,
\end{equation}
\noindent
where $|\phi_N|$ represents the Newtonian potential \citep[see,
  e.g.,][]{HuSawicki07,Cabre12}. In virial equilibrium the square of
the three-dimensional velocity dispersion is approximately equal to
the gravitational potential: ignoring the difference between the
actual gravitational potential and the Newtonian gravitational
potential\footnote{This should not be off by more than a factor $4/3$,
  which is the maximum enhancement of gravity in $f(R)$-models, and
  translates into a shift smaller than $\sqrt{4/3}-1\approx15$ per
  cent in the velocity dispersion.} we can thus get a simple estimate
of the velocity dispersion above which the Chameleon mechanism is
active. Using these assumptions, we find that modifications of gravity
are screened in halos with:

\begin{equation}
\sigma_{200} \gtrsim \sqrt{\frac{c^2 \, |\bar{f}_R(z)|}{2}}
\end{equation}
\noindent
In Figure~\ref{fig:virial}, we mark with an horizontal dotted line the
redshift dependent velocity dispersion threshold separating screened
and unscreened systems. We then consider unscreened $> 10^{12}
M_\odot$ DM haloes and we show that they follow the same scaling
relation as in Eq.~(\ref{eq:scaling}), with a different normalization
(solid line). We have thus modified \munich to rescale the assumed
virial relations of unscreened haloes to account for this mean offset
in \fr runs. We defer a more detailed study of the changes of the
virial properties of DM haloes as a function of $|\bar{f}_{R0}|$ to a
forthcoming paper (Arnold et al., in preparation).

In the following, we refrain from discussing possible changes in the
model calibrations, instead we prefer to highlight differences induced
by changes in the cosmology alone, holding all other ingredients
fixed.  As we retain the original parameter choices, this implies that
only for the \lcdm run the models are tuned to perform best when
compared to observational constraints. Nonetheless, this approach
seems best suited to assess the size of the expected effects from
modified gravity cosmologies alone.

\section{Results and Discussion}\label{sec:results}
\begin{figure}
  \centerline{ \includegraphics[width=9cm]{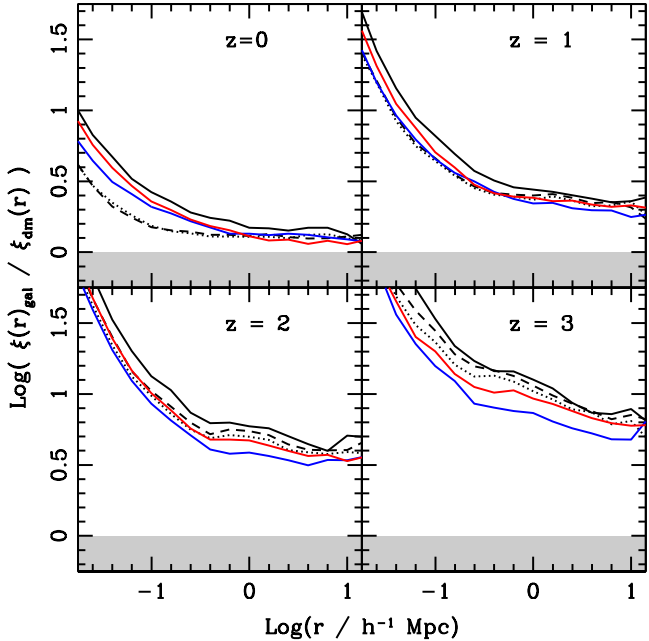} }
  \caption{Redshift evolution of the galaxy bias in different
    cosmological models. Only model galaxies with $M_\star > 10^9
    M_\odot$ have been considered. Models are labelled with the same
    line types and colours as in
    Fig.~\ref{fig:gen_sam}.}\label{fig:bias}
\end{figure}
\begin{figure*}
  \centerline{ \includegraphics[width=18cm]{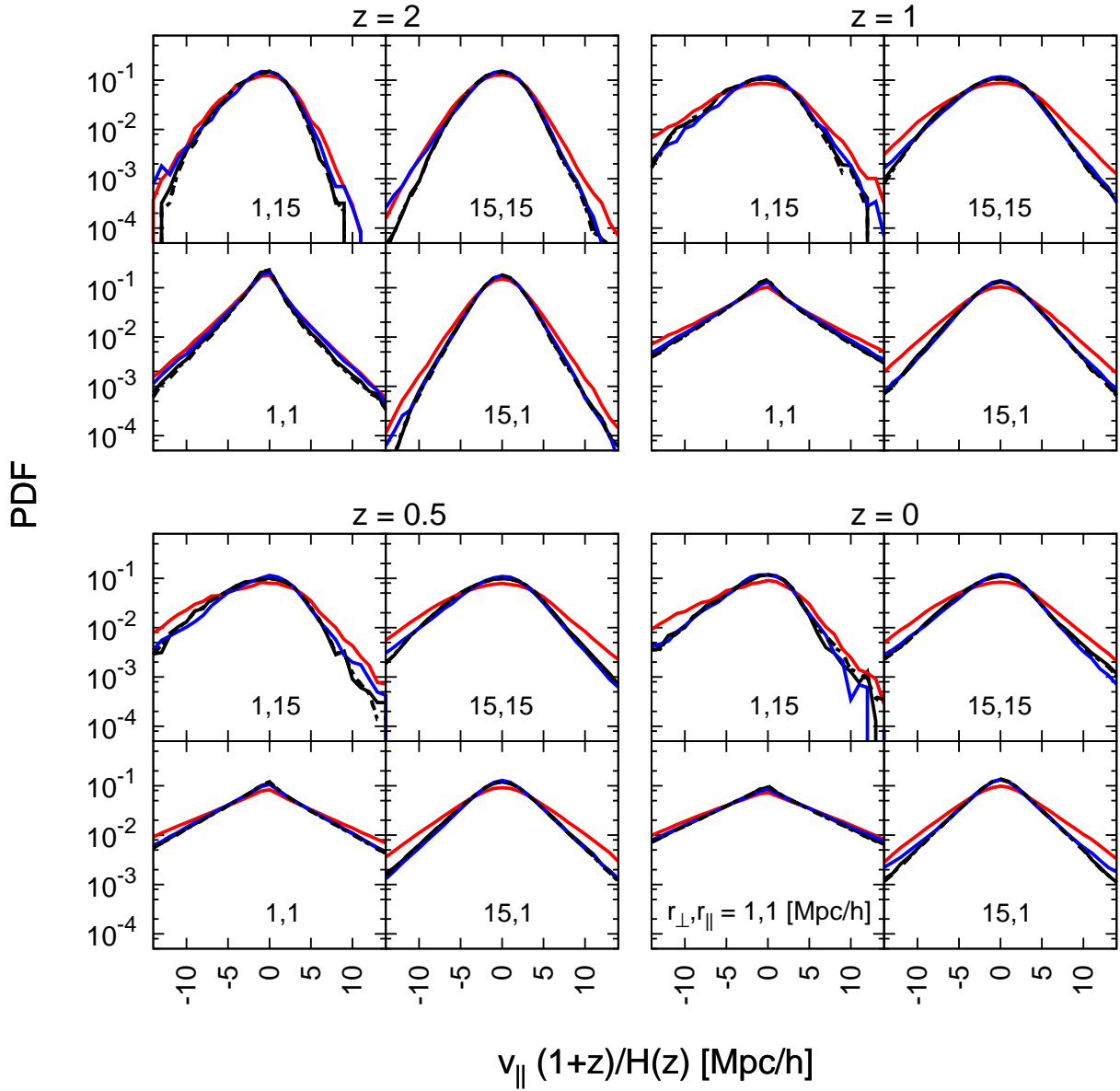} }
\caption{Pairwise galaxy velocity distribution along the line of sight
  for different models at four different redshifts. Velocities have
  been rescaled to comoving distances by the conformal Hubble function
  $\mathcal{H}=aH$. Each sub-panel represents different values of the
  galaxy separation ($r_\parallel$, $r_\perp$), parallel and
  perpendicular to the line of sight, respectively (as labelled). As
  in Fig.~\ref{fig:gen_sam}, the solid black/blue/red lines refer to
  our SAM predictions in $\Lambda$CDM, Early Dark Energy (EDE3) and
  \fr (FoR1) cosmologies, while the black dashed and dotted lines
  refer to the predictions of the \citet{DeLuciaBlaizot07} and
  \citet{Croton06} SAMs for the $\Lambda$CDM
  cosmology.}\label{fig:pairwise}
\end{figure*}

In Figure~\ref{fig:gen_sam}, we compare the redshift evolution of a
selection of galaxy properties as predicted by the different SAMs
discussed in the previous section. We choose the same quantities as in
Paper I {\it and we compare them over a common redshift range}: the
galaxy stellar mass function (upper panel) and the cosmic star
formation rate (lower panel). Whenever model predictions are compared
to observational constraints, we convolve them with an adequate
estimate of the typical observational error (e.g. a lognormal error
distribution with amplitude 0.25 and 0.3 for stellar masses and star
formation rates respectively, see \citealt{Fontanot09b}). Moreover,
the predictions of the \citet{Croton06} model have been converted from
a Salpeter to a Chabrier IMF by assuming a constant shift of 0.25 dex
in stellar mass and 0.176 dex in star formation rate. In the
following, only galaxies with $M_\star > 10^9 M_\odot$ have been
considered.

In all panels, black lines refer to the \munich prediction in the
\lcdm simulation, with solid, dashed and dotted lines referring to the
\citet{Guo11}, \citet{DeLuciaBlaizot07} and \citet{Croton06} SAMs,
respectively. We stress again that the differences in this set of
predictions are driven by the different treatment of physical
processes and varying calibration sets among the 3 models (see also
Paper I), and represent a fair estimate of the intra-SAM
variance. Predictions relative to the \citet{Guo11} model in different
cosmologies are instead shown as coloured lines: black, blue and red
refer to the $\Lambda$CDM, EDE3 and FoR1 runs, respectively. For a
full analysis of the comparison between the \lcdm and EDE models we
refer the reader to Paper I, while here we focus on the FoR1 run to
see how it complements our conclusion in Paper I. The SAM predictions
for the FoR1 run show smaller deviations relative to the \lcdm model
than a moderate EDE model (like EDE3). Moreover, these deviations do
not grow with redshift (as in the EDE3 model), but remain of small
amplitude over the whole redshift range under scrutiny. The similarity
of $z=0$ predictions between the \lcdm and \fr runs is particularly
important, since it justifies our choice of keeping the same SAM
parameters in the two runs, despite the two simulations are not
constrained to have the same matter power spectrum at $z=0$.

An interesting difference with respect to EDE models lies in the
differential effect of \fr on the stellar mass function at different
mass scales, as highlighted in the lower row, using the ratio between
the mass function in a given cosmology and the corresponding mass
function in the \lcdm run. In fact, while the EDE cosmologies predict
an enhancement of the space density at all mass scales (as a
consequence of the earlier structure formation epoch), in the FoR1 run
the space density of low-mass galaxies is systematically increased at
all redshifts, this effect being slightly larger at higher
redshifts. At the high mass end, the \fr simulation predicts a higher
space density of DM haloes than in the \lcdm run \citep[see
  e.g.][]{Li11,Zhao11}, but this effect is at most weakly reflected in
the space density of massive galaxies. We indeed find a small excess
of massive galaxies at the highest mass bin probed by our volume, but
we also find a weak deficiency around the knee of the mass
function. In the \munich model, the position of the knee is largely
determined by the `radio'-mode AGN feedback parametrization, which
assumes an explicit dependence on the virial properties of the hosting
DM halo. Given the larger virial velocities in \fr at fixed halo mass
(Figure~\ref{fig:virial}) and our choice to not modify the SAM
parameters, we expect AGN feedback to become effective at relatively
smaller halo masses, thus explaining this finding.  Nonetheless, it is
worth stressing that these effects are predicted to be smaller than
the assumed uncertainty in the stellar mass determination (and
definitely smaller than the intra-model dispersion of SAMs in \lcdm
models). It is also interesting to note that they go in the opposite
direction than required for solving the well known discrepancies
between observed and theoretically predicted mass functions \citep[see
  e.g.][]{Weinmann12}.

In Paper I, we also proposed galaxy bias\footnote{Galaxy bias has been
  defined as the ratio between the galaxy auto-correlation function
  $\xi_{\rm gal}$ and the auto-correlation function $\xi_{\rm dm}$ for
  a randomly selected subsample of DM particles corresponding to $1$
  per cent of the total particles in the cosmological box.} as a
relevant discriminant between \lcdm and EDE models. The same
conclusion does not hold in \fr models (Figure~\ref{fig:bias}): SAMs
predict similar levels of galaxy bias for both the FoR1 run and the
\lcdm box, especially at scales accessible with present and future
experiments (i.e. larger than a few Mpc).

It is well known that the redshift-space clustering of galaxies
carries an imprint of the growth of structure, providing fundamental
information about the nature of gravity \citep{Kaiser87, Zhang07,
  Guzzo08, Reyes10, Jennings12}. Standard techniques to extract this
information are traditionally based on measurements of the anisotropy
of the redshift-space correlation function
$\xi_z(r_\parallel,r_\perp)$, where $r_\parallel$ and $r_\perp$
represent the components of galaxy separation parallel and
perpendicular to the line of sight, respectively. In this work, the
volume of the simulated cosmological boxes is too small and the cosmic
variance too large to reliably apply this technique. Nonetheless, it
can still be of interest to look at the behaviour of gravity-driven
large scale coherent motions, which are a direct consequence of the
growth of structure.  Given the real-space correlation function
$\xi_r$, the corresponding $\xi_z$ is fully determined by the pairwise
galaxy velocity distribution along the line of sight
$\mathcal{P}(v_\parallel,r_\parallel,r_\perp)$, measured at each given
separation $r_\parallel$, $r_\perp$ \citep{Scoccimarro04}. In
Figure~\ref{fig:pairwise}, we show the behaviour of this distribution
in our simulations for a few fixed values of $r_\parallel$, $r_\perp$,
using galaxies with $M_\star > 10^9 M_\odot$. The velocities have been
rescaled to comving distances by the conformal Hubble function
$\mathcal{H}=aH$ so that the distribution actually represents the
statistical displacement of galaxy pairs from real to redshift
space. We adopt the convention that the pairwise velocity is negative
when galaxies are approaching each other and vice versa.  As expected,
there is a clear statistical difference between the shapes of the
distributions corresponding to the FoR1 run and all other SAMs. In
particular, \fr (red lines) predicts a larger variance in the
distribution. This discrepancy is more relevant for larger parallel
and perpendicular separations, almost vanishing on small scales
(bottom left of each panel), thus stressing the case for large surveys
as cosmological probes of gravity and dark energy.

\section{Conclusions}
\label{sec:final}

In this work, we analysed an updated version of the \munich
semi-analytic model, specifically designed to run self-consistently on
high-resolution $N$-body simulations of \fr scenarios
\citep{Puchwein13}. With respect to the \lcdm version of the code, our
modifications include the implementation of a user-defined Hubble
function and adjusted DM halo virial scalings that account for the
expected change of the mass--temperature relation (directly calibrated
on the \fr simulation of interest). This allows us to predict the
properties of galaxy populations for a \fr model and compare them with
comparable models for \lcdm and EDE cosmologies. Even though we
adopted a relatively large $|\bar{f}_{R0}|$ value (already in tension
with local test of gravity), we find that \fr has only a sutble impact
on the predicted galaxy properties, leading to deviations from \lcdm
of a size comparable to the uncertainties in the determination of
physical quantities (stellar masses and star formation rates) from
observations, and smaller than the intra-SAM variance at fixed
cosmology.

The weak constraints on cosmology coming from predicted galaxy
properties are consistent with our conclusions in Paper I, and also
with the documented response of SAMs against small variations of the
\lcdm cosmological parameters \citep[see e.g.][]{Guo11}. In order to
break some of these degeneracies, we considered the imprint of \fr on
the large scale structure assembly. We have tested the galaxy pairwise
velocity distribution along the line of sight at different
separations, and have shown the potential of a significant galaxy
sample to detect deviations from general relativity on large scales
using such statistics.
%

These results extend our previous findings for quintessence
cosmologies \citep{Fontanot12c}, and represent a step forward in a
quantitatively more accurate understanding of the relation between
cosmological scenarios and the physics of galaxy formation. This
question is indeed of fundamental importance in order to identify a
set of reliable and orthogonal observational tests able to
discriminate between the many proposed alternative cosmological
theories \citep{EuclidTheory13}. A case in point for our approach is
that in Paper I we showed that galaxy bias is a sensitive probe for
detecting an evolution of the DE equation of state (e.g. quintessence
models), while in this work we show that this probe is almost
insensitive to the effect of \fr models. On the other hand, the
analysis of the velocity distribution suggests that the anisotropy of
redshift-space correlation function and/or power spectrum as measured
from galaxy samples is a sensible probe for \fr models, but it is not
able to differentiate between \lcdm and EDE cosmologies. Overall, this
initial exploration of the coupling of semi-analytic models with full
non-linear N-body models of non-standard cosmologies shows the power
of this approach in helping to make full use of the wealth of
information expected from future wide galaxy surveys (like the EUCLID
mission, \citealt{Laureijs11}). In forthcoming work, we plan to extend
the approach further by considering other cosmological models with
alternative DE scenarios (see, e.g., \citealt{Baldi12}), as well as
considering larger box-size simulations to improve the statistical
power of the models, particularly on large scales.

\section*{Acknowledgements}
The authors thank Luigi Guzzo for enlightening discussions. FF, EP and
VS acknowledge financial support from the Klaus Tschira Foundation and
the Deutsche Forschungsgemeinschaft through Transregio 33, `The Dark
Universe'. The simulations were carried out on the `Magny' cluster of
the Heidelberger Institute f\"ur Theoretische Studien.

\bibliographystyle{mn2e}
\bibliography{fontanot}

\end{document}


%% file: DEmodels.v3.bbl
\begin{thebibliography}{}

\bibitem[\protect\citeauthoryear{{Amendola}, {Appleby}, {Bacon}, {Baker},
  {Baldi}, {Bartolo}, {Blanchard}, {Bonvin} \& et al.}{{Amendola}
  et~al.}{2012}]{EuclidTheory13}
{Amendola} L.,  {Appleby} S.,  {Bacon} D.,  {Baker} T.,  {Baldi} M.,  {Bartolo}
  N.,  {Blanchard} A.,  {Bonvin} C.,    et al. 2012, ArXiv e-prints
  (arXiv:1206.1225)

\bibitem[\protect\citeauthoryear{{Baldi}}{{Baldi}}{2012}]{Baldi12}
{Baldi} M.,  2012, \mnras, 422, 1028

\bibitem[\protect\citeauthoryear{{Boylan-Kolchin}, {Bullock} \&
  {Kaplinghat}}{{Boylan-Kolchin} et~al.}{2012}]{BoylanKolchin12}
{Boylan-Kolchin} M.,  {Bullock} J.~S.,    {Kaplinghat} M.,  2012, \mnras, 422,
  1203

\bibitem[\protect\citeauthoryear{{Cabr{\'e}}, {Vikram}, {Zhao}, {Jain} \&
  {Koyama}}{{Cabr{\'e}} et~al.}{2012}]{Cabre12}
{Cabr{\'e}} A.,  {Vikram} V.,  {Zhao} G.-B.,  {Jain} B.,    {Koyama} K.,  2012,
  \jcap, 7, 34

\bibitem[\protect\citeauthoryear{{Croton}, {Springel}, {White}, {De Lucia},
  {Frenk}, {Gao}, {Jenkins}, {Kauffmann}, {Navarro} \& {Yoshida}}{{Croton}
  et~al.}{2006}]{Croton06}
{Croton} D.~J.,  {Springel} V.,  {White} S.~D.~M.,  {De Lucia} G.,  {Frenk}
  C.~S.,  {Gao} L.,  {Jenkins} A.,  {Kauffmann} G.,  {Navarro} J.~F.,
  {Yoshida} N.,  2006, \mnras, 365, 11

\bibitem[\protect\citeauthoryear{{De Lucia} \& {Blaizot}}{{De Lucia} \&
  {Blaizot}}{2007}]{DeLuciaBlaizot07}
{De Lucia} G.,  {Blaizot} J.,  2007, \mnras, 375, 2

\bibitem[\protect\citeauthoryear{{Evrard}, {Bialek}, {Busha}, {White}, {Habib},
  {Heitmann}, {Warren}, {Rasia}, {Tormen}, {Moscardini}, {Power}, {Jenkins},
  {Gao}, {Frenk}, {Springel}, {White} \& {Diemand}}{{Evrard}
  et~al.}{2008}]{Evrard08}
{Evrard} A.~E.,  {Bialek} J.,  {Busha} M.,  {White} M.,  {Habib} S.,
  {Heitmann} K.,  {Warren} M.,  {Rasia} E.,  {Tormen} G.,  {Moscardini} L.,
  {Power} C.,  {Jenkins} A.~R.,  {Gao} L.,  {Frenk} C.~S.,  {Springel} V.,
  {White} S.~D.~M.,    {Diemand} J.,  2008, \apj, 672, 122

\bibitem[\protect\citeauthoryear{{Fontanot}, {De Lucia}, {Monaco}, {Somerville}
  \& {Santini}}{{Fontanot} et~al.}{2009}]{Fontanot09b}
{Fontanot} F.,  {De Lucia} G.,  {Monaco} P.,  {Somerville} R.~S.,    {Santini}
  P.,  2009, \mnras, 397, 1776

\bibitem[\protect\citeauthoryear{{Fontanot}, {Springel}, {Angulo} \&
  {Henriques}}{{Fontanot} et~al.}{2012}]{Fontanot12c}
{Fontanot} F.,  {Springel} V.,  {Angulo} R.~E.,    {Henriques} B.,  2012,
  \mnras, 426, 2335

\bibitem[\protect\citeauthoryear{{Grossi} \& {Springel}}{{Grossi} \&
  {Springel}}{2009}]{GrossiSpringel09}
{Grossi} M.,  {Springel} V.,  2009, \mnras, 394, 1559

\bibitem[\protect\citeauthoryear{{Guo}, {White}, {Boylan-Kolchin}, {De Lucia},
  {Kauffmann}, {Lemson}, {Li}, {Springel} \& {Weinmann}}{{Guo}
  et~al.}{2011}]{Guo11}
{Guo} Q.,  {White} S.,  {Boylan-Kolchin} M.,  {De Lucia} G.,  {Kauffmann} G.,
  {Lemson} G.,  {Li} C.,  {Springel} V.,    {Weinmann} S.,  2011, \mnras, 413,
  101

\bibitem[\protect\citeauthoryear{{Guzzo}, {Pierleoni}, {Meneux}, {Branchini},
  {Le F{\`e}vre}, {Marinoni}, {Garilli} \& {Blaizot}}{{Guzzo}
  et~al.}{2008}]{Guzzo08}
{Guzzo} L.,  {Pierleoni} M.,  {Meneux} B.,  {Branchini} E.,  {Le F{\`e}vre} O.,
   {Marinoni} C.,  {Garilli} B.,    {Blaizot} J. e.~a.,  2008, \nat, 451, 541

\bibitem[\protect\citeauthoryear{{Henriques}, {Thomas}, {Oliver} \&
  {Roseboom}}{{Henriques} et~al.}{2009}]{Henriques09}
{Henriques} B.~M.~B.,  {Thomas} P.~A.,  {Oliver} S.,    {Roseboom} I.,  2009,
  \mnras, 396, 535

\bibitem[\protect\citeauthoryear{{Henriques}, {White}, {Thomas}, {Angulo},
  {Guo}, {Lemson} \& {Springel}}{{Henriques} et~al.}{2013}]{Henriques13}
{Henriques} B.~M.~B.,  {White} S.~D.~M.,  {Thomas} P.~A.,  {Angulo} R.~E.,
  {Guo} Q.,  {Lemson} G.,    {Springel} V.,  2013, \mnras, 431, 3373

\bibitem[\protect\citeauthoryear{{Hinterbichler} \& {Khoury}}{{Hinterbichler}
  \& {Khoury}}{2010}]{HinterbichlerKhoury10}
{Hinterbichler} K.,  {Khoury} J.,  2010, Physical Review Letters, 104, 231301

\bibitem[\protect\citeauthoryear{{Hopkins}}{{Hopkins}}{2004}]{Hopkins04}
{Hopkins} A.~M.,  2004, \apj, 615, 209

\bibitem[\protect\citeauthoryear{{Hu} \& {Sawicki}}{{Hu} \&
  {Sawicki}}{2007}]{HuSawicki07}
{Hu} W.,  {Sawicki} I.,  2007, \prd, 76, 064004

\bibitem[\protect\citeauthoryear{{Jennings}, {Baugh}, {Li}, {Zhao} \&
  {Koyama}}{{Jennings} et~al.}{2012}]{Jennings12}
{Jennings} E.,  {Baugh} C.~M.,  {Li} B.,  {Zhao} G.-B.,    {Koyama} K.,  2012,
  \mnras, 425, 2128

\bibitem[\protect\citeauthoryear{{Kaiser}}{{Kaiser}}{1987}]{Kaiser87}
{Kaiser} N.,  1987, \mnras, 227, 1

\bibitem[\protect\citeauthoryear{{Khoury} \& {Weltman}}{{Khoury} \&
  {Weltman}}{2004}]{KhouryWeltman04}
{Khoury} J.,  {Weltman} A.,  2004, \prd, 69, 044026

\bibitem[\protect\citeauthoryear{{Khoury} \& {Wyman}}{{Khoury} \&
  {Wyman}}{2009}]{KhouryWyman09}
{Khoury} J.,  {Wyman} M.,  2009, \prd, 80, 064023

\bibitem[\protect\citeauthoryear{{Komatsu}, {Dunkley}, {Nolta}, {Bennett},
  {Gold}, {Hinshaw}, {Jarosik}, {Larson}, {Limon}, {Page}, {Spergel},
  {Halpern}, {Hill}, {Kogut}, {Meyer}, {Tucker}, {Weiland}, {Wollack} \&
  {Wright}}{{Komatsu} et~al.}{2009}]{Komatsu09}
{Komatsu} E.,  {Dunkley} J.,  {Nolta} M.~R.,  {Bennett} C.~L.,  {Gold} B.,
  {Hinshaw} G.,  {Jarosik} N.,  {Larson} D.,  {Limon} M.,  {Page} L.,
  {Spergel} D.~N.,  {Halpern} M.,  {Hill} R.~S.,  {Kogut} A.,  {Meyer} S.~S.,
  {Tucker} G.~S.,  {Weiland} J.~L.,  {Wollack} E.,    {Wright} E.~L.,  2009,
  \apjs, 180, 330

\bibitem[\protect\citeauthoryear{{Laureijs}, {Amiaux}, {Arduini},
  {Augu{\`e}res}, {Brinchmann}, {Cole}, {Cropper}, {Dabin}, {Duvet}, {Ealet} \&
  et al.}{{Laureijs} et~al.}{2011}]{Laureijs11}
{Laureijs} R.,  {Amiaux} J.,  {Arduini} S.,  {Augu{\`e}res} J.~.,  {Brinchmann}
  J.,  {Cole} R.,  {Cropper} M.,  {Dabin} C.,  {Duvet} L.,  {Ealet} A.,    et
  al. 2011, ArXiv e-prints (arXiv:1110.3193)

\bibitem[\protect\citeauthoryear{{Li}, {Zhao} \& {Koyama}}{{Li}
  et~al.}{2012}]{LiZhaoKoyama12}
{Li} B.,  {Zhao} G.-B.,    {Koyama} K.,  2012, \mnras, 421, 3481

\bibitem[\protect\citeauthoryear{{Li} \& {Hu}}{{Li} \& {Hu}}{2011}]{Li11}
{Li} Y.,  {Hu} W.,  2011, \prd, 84, 084033

\bibitem[\protect\citeauthoryear{{Macci{\`o}}, {Kang}, {Fontanot},
  {Somerville}, {Koposov} \& {Monaco}}{{Macci{\`o}} et~al.}{2010}]{Maccio10}
{Macci{\`o}} A.~V.,  {Kang} X.,  {Fontanot} F.,  {Somerville} R.~S.,  {Koposov}
  S.,    {Monaco} P.,  2010, \mnras, 402, 1995

\bibitem[\protect\citeauthoryear{{McCarthy}, {Bower} \& {Balogh}}{{McCarthy}
  et~al.}{2007}]{McCarthy07}
{McCarthy} I.~G.,  {Bower} R.~G.,    {Balogh} M.~L.,  2007, \mnras, 377, 1457

\bibitem[\protect\citeauthoryear{{Planck Collaboration}, {Ade}, {Aghanim},
  {Armitage-Caplan}, {Arnaud}, {Ashdown}, {Atrio-Barandela}, {Aumont},
  {Baccigalupi}, {Banday} \& et al.}{{Planck Collaboration}
  et~al.}{2013}]{Planck_cosmpar}
{Planck Collaboration} {Ade} P.~A.~R.,  {Aghanim} N.,  {Armitage-Caplan} C.,
  {Arnaud} M.,  {Ashdown} M.,  {Atrio-Barandela} F.,  {Aumont} J.,
  {Baccigalupi} C.,  {Banday} A.~J.,    et al. 2013, ArXiv e-prints
  (arXiv:1303.5076)

\bibitem[\protect\citeauthoryear{{Puchwein}, {Baldi} \& {Springel}}{{Puchwein}
  et~al.}{2013}]{Puchwein13}
{Puchwein} E.,  {Baldi} M.,    {Springel} V.,  2013, ArXiv e-prints
  (arXiv:1305.2418)

\bibitem[\protect\citeauthoryear{{Reyes}, {Mandelbaum}, {Seljak}, {Baldauf},
  {Gunn}, {Lombriser} \& {Smith}}{{Reyes} et~al.}{2010}]{Reyes10}
{Reyes} R.,  {Mandelbaum} R.,  {Seljak} U.,  {Baldauf} T.,  {Gunn} J.~E.,
  {Lombriser} L.,    {Smith} R.~E.,  2010, \nat, 464, 256

\bibitem[\protect\citeauthoryear{{Schmidt}}{{Schmidt}}{2009}]{Schmidt09b}
{Schmidt} F.,  2009, \prd, 80, 043001

\bibitem[\protect\citeauthoryear{{Scoccimarro}}{{Scoccimarro}}{2004}]{Scoccima%
rro04}
{Scoccimarro} R.,  2004, \prd, 70, 083007

\bibitem[\protect\citeauthoryear{{Springel}}{{Springel}}{2005}]{Springel05c}
{Springel} V.,  2005, \mnras, 364, 1105

\bibitem[\protect\citeauthoryear{{Springel}, {White}, {Jenkins}, {Frenk},
  {Yoshida}, {Gao}, {Navarro}, {Thacker}, {Croton}, {Helly}, {Peacock}, {Cole},
  {Thomas}, {Couchman}, {Evrard}, {Colberg} \& {Pearce}}{{Springel}
  et~al.}{2005}]{Springel05}
{Springel} V.,  {White} S.~D.~M.,  {Jenkins} A.,  {Frenk} C.~S.,  {Yoshida} N.,
   {Gao} L.,  {Navarro} J.,  {Thacker} R.,  {Croton} D.,  {Helly} J.,
  {Peacock} J.~A.,  {Cole} S.,  {Thomas} P.,  {Couchman} H.,  {Evrard} A.,
  {Colberg} J.,    {Pearce} F.,  2005, \nat, 435, 629

\bibitem[\protect\citeauthoryear{{Springel}, {White}, {Tormen} \&
  {Kauffmann}}{{Springel} et~al.}{2001}]{Springel01}
{Springel} V.,  {White} S.~D.~M.,  {Tormen} G.,    {Kauffmann} G.,  2001,
  \mnras, 328, 726

\bibitem[\protect\citeauthoryear{{Vainshtein}}{{Vainshtein}}{1972}]{Vainshtein%
72}
{Vainshtein} A.~I.,  1972, Physics Letters B, 39, 393

\bibitem[\protect\citeauthoryear{{Weinmann}, {Kauffmann}, {van den Bosch},
  {Pasquali}, {McIntosh}, {Mo}, {Yang} \& {Guo}}{{Weinmann}
  et~al.}{2009}]{Weinmann09}
{Weinmann} S.~M.,  {Kauffmann} G.,  {van den Bosch} F.~C.,  {Pasquali} A.,
  {McIntosh} D.~H.,  {Mo} H.,  {Yang} X.,    {Guo} Y.,  2009, \mnras, 394, 1213

\bibitem[\protect\citeauthoryear{{Weinmann}, {Pasquali}, {Oppenheimer},
  {Finlator}, {Mendel}, {Crain} \& {Macci{\`o}}}{{Weinmann}
  et~al.}{2012}]{Weinmann12}
{Weinmann} S.~M.,  {Pasquali} A.,  {Oppenheimer} B.~D.,  {Finlator} K.,
  {Mendel} J.~T.,  {Crain} R.~A.,    {Macci{\`o}} A.~V.,  2012, \mnras, 426,
  2797

\bibitem[\protect\citeauthoryear{{Zhang}, {Liguori}, {Bean} \&
  {Dodelson}}{{Zhang} et~al.}{2007}]{Zhang07}
{Zhang} P.,  {Liguori} M.,  {Bean} R.,    {Dodelson} S.,  2007, Physical Review
  Letters, 99, 141302

\bibitem[\protect\citeauthoryear{{Zhao}, {Li} \& {Koyama}}{{Zhao}
  et~al.}{2011}]{Zhao11}
{Zhao} G.-B.,  {Li} B.,    {Koyama} K.,  2011, \prd, 83, 044007

\end{thebibliography}
